
\baselineskip=16pt

\noindent
{\bf ON A POSSIBLE ALGEBRA MORPHISM OF U$_q$[OSP(1/2N)] ONTO
  THE DEFORMED

\noindent
OSCILLATOR ALGEBRA W$_q$(N)}

\vskip 32pt

\noindent
T. D. Palev* and N. I. Stoilova*

\noindent
International Centre for Theoretical Physics, 34100 Trieste, Italy
\vskip 12pt

\footnote{*}{Permanent address: Institute for Nuclear Research and
Nuclear Energy, 1784 Sofia, Bulgaria; E-mail address:
palev@bgearn.bitnet}

\vskip 32pt

\leftskip 32pt

\noindent
{\bf Abstract}. We formulate a conjecture, stating that the algebra of
 $n$ pairs  of deformed Bose creation and annihilation operators is a
factor-algebra of $U_q[osp(1/2n)]$, considered as a Hopf algebra, and
 prove it for $n=2$ case. To this end we show that  for any value of $q$
$U_q[osp(1/4)]$ can be viewed as
a superalgebra, freely generated by two pairs $B_1^\pm$, $B_2^\pm$ of
deformed para-Bose operators. We write down all Hopf algebra relations,
 an analogue of the  Cartan-Weyl
 basis, the "commutation" relations between the generators and a basis in
$U_q[osp(1/2n)]$ entirely in terms of  $B_1^\pm$, $B_2^\pm$.


\vskip 32pt
\leftskip 12pt

{\bf I. Introduction} \vskip 12pt

One way to describe completely a given simple Lie (super)algebra $A$
is in terms of its Chevalley generators. These generators are especially
appropriate for a quantization of $A$, i.e.,  for a deformation of the
universal enveloping algebra $U[A]$ of $A$ to  a new associative
algebra $U_q[A]$ in such a way that $U_q[A]$ remains a Hopf algebra.

Another possible way to define $A$ and  $U[A]$ was outlined in Ref.1. It
is based on the concept of creation and annihilation operators (CAO's)
of the simple Lie (super)algebra $A$ under consideration. Contrary to the
Chevalley generators, the  creation and annihilation operators of some
algebras have direct physical significance. The CAO's $F_1^\pm,F_2^\pm,
\ldots,F_n^\pm$ of the orthogonal Lie algebra $B_n\equiv so(2n+1)$ are
known in quantum field theory as para-Fermi operators [2,3]; in a
particular representation of  $B_n$ they become usual Fermi operators.
Similarly, the CAO's  $B_1^\pm,B_2^\pm,
\ldots,B_n^\pm$ of the orthosymplectic Lie
superalgebra  $osp(1/2n)$ are the para-Bose operators [5,6]; in the
representation, corresponding to an order of the statistics $p=1$ they
reduce to Bose creation and annihilation operators
 $b_1^\pm,b_2^\pm,
\ldots,b_n^\pm$.

Clearly any deformation of $U[osp(1/2n)]$ will lead to a deformation of
 $B_1^\pm,B_2^\pm,\ldots,B_n^\pm$
and consequently to a deformation of the Bose operators
 $b_1^\pm,b_2^\pm,\ldots,b_n^\pm$.
In this relation we wish to rise   and discuss two questions.

1. Do the deformed CAO's $B_1^\pm,B_2^\pm,\ldots,B_n^\pm$ of $osp(1/2n)$
define entirely  $U_q[osp(1/2n)]$ (as this is the case with the deformed
Chevalley generators)?

2. Is there any relation between the deformation of the Bose operators,
obtained in this way, and the other known approaches to deform the Bose
operators [7-11], which are unrelated to any Hopf algebra structure?

At present we do not know the answers neither to the first nor to the
second question. There are good evidences, however, that the answer to both
questions is positive and in particular that the Fock space representation
of the $q-$deformed CAO's
 $B_1^\pm,B_2^\pm,\ldots,B_n^\pm$
coincides with the deformed Bose operators as defined in Refs.8-10.

In order to formulate our conjecture more precisely let $W_q(n)$ be the
deformed Weyl (or oscillator) algebra as defined by Hayashi [12].
The oscillator algebra $W_q(n)$ is an
associative algebra with unity, free generators
$b_i^\pm, k_i^\pm, i=1,\ldots,n$
and the relations ($i,j=1,\ldots,n$)
$$k_i^{-1}k_i=k_ik_i^{-1}=1$$
$$k_ib_i^\pm=q^{\pm 1} b_i^\pm k_i$$
$$b_i^-b_i^+-q^2b_i^+b_i^-=k_i^{-2} \eqno(1.2)$$
$$b_i^-b_i^+-q^{-2}b_i^+b_i^-=k_i^2$$
$$a_ia_j=a_ja_i,  \hskip 12pt  i\neq j,$$
where $a_i=b_i^\pm, k_i^{\pm 1}$.

To turn $W_q(n)$ into a superalgebra (  {\bf Z}$_2$-graded algebra) we set

 $$deg(b_i^\pm)=1 \in {\bf Z}_2 , \hskip 12pt deg(k_i^{\pm 1})=0 \in {\bf Z}_2,
\hskip 12pt  i=1,\ldots,n. \eqno(1.3)$$

In the Fock representation of $W_q(n)$ , namely when $k_i=q^{N_i}$,
$b_i^\pm$ are
the deformed $q$-bosons [8-10] and $N_i$ is the $i$th boson number operator.
With respect to the grading following from (1.3) $W_q(n)$ is an
infinite-dimensional associative superalgebra. It is neither a Hopf algebra
nor even a coalgebra. Our conjecture
is the following one.

CONJECTURE.{\it The deformed Weyl superalgebra $W_q(n)$ is a factor algebra
of a deformed universal enveloping algebra $U_q[osp(1/2n)]$.}

This conjecture holds in the nondeformed case [6]. In Sec.II we recall the
idea of the proof. At $q \not= 1$ the conjecture has been proved so far for
$n=1$ [13]. Here we prove it for $n=2$. To this end in Sec.III we deform
$U[osp(1/4)]$ in terms of generators, which are in fact deformed para-Bose
operators.

The case $n=2$ was considered in Ref.14 in relation to the "supersingleton"
Fock representation of $osp(1/4)  $ and its "singleton" [15] structure.
The authors have studied in details the quantum deformations of $sp(4)$ at
$q$ being root of unity using deformed CAO's. We have been informed they
have similar results to ours also for the deformed  $osp(1/4)$ [16].

If the conjecture turns true, one can use the Hopf algebra structure in order
to construct new representations of $U_q[osp(1/2n)]$ or of any of
 its subalgebras
beginning  with some known representations of it and in particular
with the representation  ${\rho}_F$ in the Fock space of the deformed Bose
operators $F_q(n) \equiv$ \hfill\break $Fock_q(n)$. To this end one can
 use the comultiplication $\Delta$.
 For instance the maps
$$\Delta^{(2)}=(\rho_F \otimes \rho_F ) \circ \Delta :
U_q[osp(1/2n)] \rightarrow End(F_q(n) \otimes F_q(n))  $$
and
$$ \Delta^{(3)}=(\rho_F \otimes \rho_F \otimes \rho_F ) \circ
[(id \otimes  \Delta) \circ \Delta ] :
U_q[osp(1/2n)] \rightarrow End(F_q(n) \otimes F_q(n)  \otimes F_q(n)  )
\eqno(1.4)  $$
define representations of $U_q[osp(1/2n)]$ in $F_q(n) \otimes F_q(n)$  and
$F_q(n) \otimes F_q(n) \otimes F_q(n)   $, respectively .

Throughout we use the following abbreviations and notation:

\vskip 6pt
LS (LS's)--- Lie superalgebra (Lie superalgebras),

$lin.env.\{X\}$ --- the linear envelope of $X$,

${\bf Z}$ --- all integers,

${\bf Z}_+$ --- all nonnegative integers,

${\bf Z}_2 = (0,1)$ --- the ring of all integers modulo 2,

$[A,B]=AB-BA, \hskip 6pt \{A,B\}=AB+BA$,

$<A,B>=AB-(-1)^{deg(A)deg(B)}BA$,

$[A,B]_{q^n}=AB-q^nBA, \hskip 12pt \{A,B\}_{q^n}=AB+q^nBA $.

\vfill 24pt

{\bf II. The Nondeformed Case}
\vskip 12pt

Let $ Free(n)  $ be the associative superalgebra with unity,
free generators
$ \quad B_1^\pm ,\quad B_2^\pm,\quad \ldots,\quad B_n^\pm $, \hfill\break
 $deg(B_i^\pm)=~1$ and the relations
($\xi, \eta, \epsilon = \pm$ or $\pm 1$, $i,j,k=1,2,\ldots ,n$)
$$[\{B_i^\xi,B_j^\eta \},B_k^\epsilon]=(\epsilon - \xi)\delta_{ik}
B_j^\eta + (\epsilon - \eta) \delta_{jk}B_i^\xi. \eqno(2.1)$$
Consider the subspace
$$B(0/n)=lin.env.\big\{\{B_i^\xi,B_j^\eta \},B_k^\epsilon \mid
\xi, \eta, \epsilon = \pm,i,j,k=1,2,\ldots ,n \big\} \eqno(2.2)$$
and define a supercommutator on it
$$<A,B>=AB-(-1)^{deg(A)deg(B)}BA, \hskip 6pt A,B \in B(0/n).\eqno(2.3)$$

PROPOSITION 1 [6]. {\it $B(0/n)$ is a Lie superalgebra isomorphic to the
orthosymplectic LS $osp(1/2n)$ with  an even subalgebra

$$sp(2n)=lin.env.\big\{\{B_i^\xi,B_j^\eta \} \mid
\xi, \eta, = \pm,i,j=1,2,\ldots ,n \big\}. \eqno(2.4)$$ }

PROPOSITION 2. {\it The associative superalgebra $Free(n)$ is (isomorphic
to) the universal enveloping algebra of $osp(1/2n)$,
$$Free(n)=U[osp(1/2n)]. \eqno(2.5)$$}

{}From these propositions one concludes  that $U[osp(1/2n)]$ is generated from
$B_1^\pm, B_2^\pm,\ldots,B_n^\pm \in osp(1/2n)$. We point out that these $2n$
generators are very different from the Chevalley generators of the same
 algebra.
The operators $B_1^\pm, B_2^\pm,\ldots,B_n^\pm$ were introduced by Green [2]
 as a possible generalization of the Bose statistics and are called
 para-Bose operators. Propositions 1 and 2 indicate that the
 representation theory of the para-Bose statistics is simply another name
 for the representation theory of the orthosymplectic Lie superalgebra.

Let $b_1^\pm, b_2^\pm,\ldots,b_n^\pm$ be Bose creation and annihilation
 operators  and let $W(n)$ be the corresponding Weyl superalgebra,
 i.e., the set of all
polynomials of $b_i^\pm$, considered as odd variables. It is straightforward
to check that the Bose operators satisfy the para-Bose relations (2.1):

$$[\{b_i^\xi,b_j^\eta \},b_k^\epsilon]=(\epsilon - \xi)\delta_{ik}
b_j^\eta + (\epsilon - \eta) \delta_{jk}b_i^\xi. \eqno(2.6)$$
This shows that  the conjecture holds in the nondeformed case:

PROPOSITION 3.{\it The Weyl superalgebra $W(n)$ is a factor-algebra of
$U[osp(1/2n)]$.}

Consequently any representation of $W(n)$ and in particular its Fock
representation is a representation of $U[osp(1/2n)]$. Hence one has

PROPOSITION 4. {\it The linear map $\rho$ defined by the replacement
$B_i^\pm \rightarrow b_i^\pm, i=1,\ldots,n$ is a morphism of
 $U[osp(1/2n)]$ onto $W(n)$.}

The above proposition is in the origin of the so called ladder
(or oscillator) representations . From (2.4) and proposition 4 one
obtaines

$$sp(2n)=lin.env.\big\{\{b_i^\xi,b_j^\eta \} \mid
\xi, \eta, = \pm,i,j=1,2,\ldots ,n \big\}. \eqno(2.7)$$
Similarly
$$gl(n)=lin.env.\big\{\{b_i^-,b_j^+ \} \mid
i,j=1,2,\ldots ,n \big\}. \eqno(2.8)$$

\vskip 24pt
{\bf III. The Superalgebra $U_q[osp(1/4)]$}
\vskip 12pt

For a quantization of  $U_q[osp(1/4)]$ in terms of its Chevalley generators
 see Refs.17, 18. Here we proceed in a different way, which will make it
 easier to prove the conjecture for $n=2$  and is of independent interest.

Let $ Free_q(B_1^\pm,B_2^\pm,K_1^{\pm 1},K_2^{\pm 1})$ be the
 associative algebra with unity, free generators
$B_1^\pm$, $B_2^\pm $,  $K_1^{\pm1}$,  $K_2^{\pm1}$
and the relations ($\xi, \eta = \pm$ or $\pm 1$)

$$K_iK_i^{-1}=K_i^{-1}K_i=1, \hskip 6pt K_1K_2=K_2K_1,
 \hskip 6pt i=1,2, \eqno(3.1)$$

$$K_i^\xi B_i^\eta=q^{\xi \eta}B_i^\eta K_i^\xi, \hskip 6pt
K_i^\xi B_j^\eta=B_j^\eta K_i^\xi, \hskip 6pt
 i\not= j =1,2 ,\eqno(3.2)$$

$$ \{B_i^- , B_i^+ \}={{qK_i^2 - q^{-1}K_i^{-2}}\over {q-q^{-1}}}
, \hskip 6pt i=1,2, \eqno(3.3)$$

$$[\{B_1^\xi , B_2^\eta \}_{q^{-2\xi \eta}},
yB_1^\epsilon]_{q^{\eta (\epsilon - \xi)}}
={1\over 2}(\epsilon - \xi)q^{-\eta(\xi +1)}(1+q^{-2\xi \eta})
B_2^\eta K_1^{-2\eta}, \eqno(3.4)  $$

$$[\{B_1^\xi , B_2^\eta \}_{q^{-2\xi \eta}},
B_2^\epsilon]_{q^{\xi(\eta- \epsilon)}}
={1\over 2}(\epsilon -\eta) (1+q^{2\xi})B_1^\xi K_2^{2\xi}. \eqno(3.5)  $$
To turn  $ Free_q(B_1^\pm,B_2^\pm,K_1^{\pm 1},K_2^{\pm 1})$
  into an associative superalgebra we set
$$deg(B_i^\pm)=1 \in {\bf Z}_2, \hskip 6pt deg(K_i^{\pm 1})=0 \in
{\bf Z}_2, \hskip 6pt  i=1,2. \eqno(3.6)$$

PROPOSITION 5.{\it $ Free_q(B_1^\pm,B_2^\pm,K_1^{\pm 1},K_2^{\pm 1})$ is a
 Hopf  superalgebra with a comultiplication $\Delta$, a counit
$\epsilon$ and an antipode $S$ as follows:}

$$ \Delta (B_1^+)=q^{-1/2}B_1^+\otimes K_1K_2^{-2} +
q^{-3/2}K_1^{-1}K_2^{-2} \otimes B_1^+ +
(q-q^{-1})q^{-1/2}B_2^+ K_1^{-1} K_2^{-1} \otimes
\{B_2^-, B_1^+\}_{q^-2}K_2^{-1}, $$

$$ \Delta (B_1^-)=q^{3/2}B_1^- \otimes K_1K_2^2
+q^{1/2}K_1^{-1}K_2^2 \otimes B_1^-
+(q^{-1}-q)q^{1/2}\{B_1^-,B_2^+\}_{q^2}K_2 \otimes B_2^- K_1K_2, $$

$$ \Delta (B_2^\xi)=q^{1/2}B_2^\xi \otimes K_2 +
q^{-1/2}K_2^{-1} \otimes B_2^\xi, \hskip 6pt \xi =\pm,$$

$$\Delta (K_i^\xi)=q^{\xi/2}K_i^\xi \otimes K_i^\xi,
\hskip 6pt  i=1,2, \hskip 6pt \xi=\pm 1, \eqno(3.7)  $$

$$ \epsilon(B_i^\xi)=0, \hskip 6pt  \epsilon(K_i^\xi)=q^{-\xi/2},
\hskip 6pt i=1,2, \hskip 6pt \xi=\pm\ or\ \pm 1,  $$

$$ S(K_i^\xi)=q^{-\xi}K_i^{-\xi}, \hskip 6pt i=1,2, \hskip 6pt
\xi=\pm 1,  $$

$$S(B_1^+)=(q^7-q^5)\{B_2^-,B_1^+\}_{q^{-2}}B_2^+K_2^2 -
q^7B_1^+ K_2^4,  $$

$$S(B_1^-)=(q^{-7}-q^{-5})B_2^-\{B_1^-,B_2^+\}_{q^2}K_2^{-2}
-q^{-7}B_1^-K_2^{-4},  $$

$$ S(B_2^\xi)=-q^\xi B_2^\xi, \hskip 6pt \xi=\pm\ or\  \pm 1.   $$

PROPOSITION  6.{\it The associative superalgebra
$\quad Free_q(B_1^\pm,B_2^\pm,K_1^{\pm 1},K_2^{\pm 1}) \quad$
 is a deformation of \hfill\break $U[osp(1/4)]$. }

For a proof set $K_i=q^{H_i-1/2}$ . Then as $q \rightarrow 1$ the relations
(3.1)--(3.5) reduce to the para-Bose relations (2.1). In particular
$H_i= {1\over 2}\{B_i^-,B_i^+\}. $

In terms of the generators  $ B_1^\pm,B_2^\pm,K_1^{\pm 1},K_2^{\pm 1} $
one can construct a $q$-analog of the Cartan-Weyl basis. For all
 values of $q$ it is given with
 the following 14 generators:

$$K_i, \hskip 6pt B_i^\pm, \hskip 6pt (B_i^\pm)^2, \hskip 6pt
\{B_1^\xi,B_2^\eta\}_{q^{-2\xi \eta}}, \hskip 6pt i=1,2, \hskip 6pt
\xi, \eta=\pm\ or \pm 1.  \eqno(3.8) $$

PROPOSITION 7. {\it All ordered monomials $(n_i,m_i \in {\bf Z}_+, \hskip 3pt
p_i \in  {\bf Z})  $

$$(B_2^+)^{n_1} \{B_1^+,B_2^+\}_{q^{-2}}^{n_2}(B_1^+)^{n_3}
\{B_2^-,B_1^+\}_{q^{-2}}^{n_4} \{B_1^-,B_2^+\}_{q^2}^{m_1}
(B_1^-)^{m_2} \{B_1^-,B_2^-\}_{q^{-2}}^{m_3}(B_2^-)^{m_4}
(K_1)^{p_1}(K_2)^{p_2} \eqno(3.9)  $$
constitute a basis in $U_q[osp(1/4)]$}.

This is a $q$-deformed version of the Poincare-Birkhoff-Witt theorem.
 The proof follows from eqs. (3.1)--(3.5) and the relations following from
 them, namely

$$[\{B_2^-,B_1^+\}_{q^{-2}},\{B_1^+,B_2^+\}_{q^{-2}}]_{q^{-2}}=
(1+q^{-2})^2(B_1^+)^2K_2^2,   $$

$$[\{B_2^-,B_1^+\}_{q^{-2}},\{B_1^-,B_2^-\}_{q^{-2}}]_{q^{2}}=
-(1+q^{-2})(1+q^2)(B_2^-)^2K_1^2,   $$

$$[\{B_2^-,B_1^+\}_{q^{-2}},\{B_1^-,B_2^+\}_{q^{2}}]=
{q+q^{-1}\over q-q^{-1}}(K_1^2K_2^{-2}-K_1^{-2}K_2^2), $$

$$[\{B_1^-,B_2^-\}_{q^{-2}},\{B_1^-,B_2^+\}_{q^{2}}]_{q^2}=
(1+q^{-2})(1+q^2)(B_1^-)^2K_2^{-2}, \eqno(3.10)     $$

$$[\{B_1^+,B_2^+\}_{q^{-2}},\{B_1^-,B_2^+\}_{q^{2}}]_{q^{-2}}=
-(1+q^{-2})^2(B_2^+)^2K_1^{-2},   $$

$$[\{B_1^+,B_2^+\}_{q^{-2}},\{B_1^-,B_2^-\}_{q^{-2}}]=
{1+q^{-2}\over 1-q^{2}}(q^2K_1^2K_2^2-q^{-2}K_1^{-2}K_2^{-2}).    $$

Observe the very interesting situation that appears at $q=\pm i$
- a case which is
not considered in terms of the Chevalley basis [17, 18]. For these values of
$q$ the right hand sides of all equations (3.10) vanish. This particular case
deserves further investigations.

\vskip 24pt
{\bf IV. Proof of the Conjecture for n=2}
\vskip 12pt

PROPOSITION 8.{\it The Weyl  superalgebra $W_q(2)$ generated  by
the deformed Bose operators $ b_1^\pm,\  k_1^{\pm 1} $,
$ b_2^\pm,\ k_2^{\pm 1}$ is a factor-algebra of
$ U_q[osp(1/4)]$.}

The proof is an immediate consequence of the observation that the deformed
 Bose operators  $ b_1^\pm$, $ k_1^{\pm 1} $,
$ b_2^\pm $, $ k_2^{\pm 1}$ satisfy the defining relations for
$ U_q[osp(1/4)]$. More precisely, eqs. (3.1)--(3.5) remain valid after the
 replacement $B_i^\pm \rightarrow b_i^\pm  $, $K_i^{\pm 1}
 \rightarrow k_i^{\pm 1} $.

Consider the representation of  $W_q(2)$ in the Fock space $Fock_q(2)$ [8-10].
Then using proposition 8 and eq.(1.4) we can write a representation of
$ U_q[osp(1/4)]$ in $Fock_q(2) \otimes Fock_q(2) $:

$$\Delta^{(2)}(B_1^+)=b_1^+ \otimes q^{N_1-2N_2-1/2} + q^{-N_1-2N_2-3/2}
\otimes b_1^+ +(q^{1/2}-q^{-7/2})b_2^+q^{-N_1-N_2} \otimes
b_1^+b_2^-q^{-N_2}, $$
$$ \Delta^{(2)}(B_1^-)=b_1^-  \otimes q^{N_1+2N_2+3/2} +
q^{-N_1+2N_2+1/2} \otimes b_1^- +(q^{-1/2}-q^{7/2})b_1^-b_2^+q^{N_2}
\otimes b_2^-q^{N_1+N_2},  $$

$$\Delta^{(2)}(B_2^\xi)=b_2^\xi \otimes q^{N_2+1/2}+q^{-N_2-1/2}
\otimes b_2^\xi, \hskip 6pt \xi=\pm, \eqno(4.1)  $$

$$\Delta^{(2)}(K_i^\xi)=q^{\xi(N_i+1/2)} \otimes q^{\xi N_i},
\hskip 6pt i=1,2, \ \xi=\pm 1.   $$

The operator $\Delta^{(2)} $  defines a morphism of $U_q[osp(1/4)]  $
into $W_q(2)\otimes W_q(2)$. In all essential points (as far as the
representations of  $U_q[osp(1/4)]  $ or of any of its subalgebras are
concerned)  $\Delta^{(2)} $ is a good substitute for a comultiplication in
the Weyl algebra $W_q(2)$. The operator  $\Delta^{(2)} $  however does
not satisfy the requirements for a comultiplication in  $W_q(2)$.
In fact it is impossible to define a comultiplication in the Weyl algebra
 even in the nondeformed case.

Equations (4.1) indicate how to construct new representations of the
 superalgebra  $U_q[osp(1/4)] $ using its oscillator representation, i.e.,
the Fock space representation of its factor-algebra  $W_q(2)$. If the
 conjecture turns true, then the same approach can be applied for any
$U_q[osp(1/2n)]$. Certainly, instead of the oscillator representation
one can use any other representation. The point is however that other
representations are at present unknown (contrary to the class of quantum
superalgebras  $U_q[gl(n/1)]$ [19]). This statement holds even for the
 ordinary,  the nondeformed case and even for $osp(1/4) $. All
 finite-dimensional representations of the orthosymplectic LS's
 $osp(2m+1/2n) $  are
completely classified [4]. However (apart
  from   $osp(1/2) $ [20] and $osp(3/2) $ [21]) explicit
expressions for the transformations of the finite-dimensional
irreducible  $osp(2m+1/2n) $ modules are not available.

\vskip 24pt
{\bf Ackowledgments}
\vskip 12pt
We are thankful to Prof. Abdus Salam, the International Atomic Energy
 Agency and UNESCO for the kind hospitality at the International
Center for TheoreticalPhysics, where part of the present research was
 completed. We are thankful also to the Committee of Science of Bulgaria
 for the partial support through  contract $\Phi - 215$.

\vskip 24pt
{\bf References}
\vskip 12pt
\noindent
1. Palev,T.D., J.Math.Phys. {\bf 21}, 1293 (1980).

\noindent
2. Green,H.S., Phys.Rev. {\bf 90} , 270 (1953).

\noindent
3.  Kamefuchi,S. and Takahashi,Y., Nucl.Phys.{\bf 36}, 177 (1960);
Ryan,C. and Sudarshan,E.C.G., Nucl.Phys.

\hskip -6pt {\bf 47},207 (1963).

\noindent
4. Kac,V.G., Lect.Notes Math. {\bf 626}, 597 (1978).

\noindent
5. Omote,M., Ohnuki,Y. and Kamefuchi,S.,Prog.Theor.Phys. {\bf 56},
1948 (1976).

\noindent
6. Ganchev,A. and Palev,T.D.,J.Math.Phys.{\bf 21}, 797 (1980).

\noindent
7. Pusz,W. and Woronowicz,S.L.,Rep.Math.Phys. {\bf 27}, 231 (1989).

\noindent
8. Biedenharn, L.C.,J.Phys. A {\bf 22}, L873 (1989).

\noindent
9. Macfarlane, A.J.,J.Phys. A {\bf 22}, 4581 (1989).

\noindent
10. Sun,C.P. and Fu, H.C.,J.Phys. A {\bf 22}, L983 (1989).

\noindent
11. Hadjiivanov, L.K., Paunov, R.R. and Todorov,I.T.,
J.Math.Phys., {\bf 33}, 1379 (1992).

\noindent
12. Hayashi, T., Comm.Math.Phys. {\bf 127}, 129 (1990).

\noindent
13. Florearini, R. and Vinet, L., Preprint UCLA/90/TEP/30;
    J.Phys. A {\bf 23}, L1019 (1990).

  \hskip -2pt
 Celeghini, E., Palev, T.D. and Tarlini,  M., Preprint
 YITP/K-879, Kyoto (1990); Mod.Phys.Lett. B

  \hskip -2pt
{\bf 5}, 187 (1991).

\noindent
14. Flato, M.,  Hadjiivanov, L.K. and Todorov, I.T., Quantum
Deformations of Singletons and of Free

\hskip -2pt Zero-mass Fields,
to appear in Foundations of Physics, the volume dedicatet to A. O. Barut.

\noindent
15. Flato, M. and and Fronsdal, C., Singletons: Fundamental Gauge Theory,
in: {\bf Topological and

\hskip -2pt Geometrical Methods in Field Theory},
Symposium in Espoo, Finland 1986, Eds. J. Hietarinta,

\hskip -2pt J. Westerholm
(World Scientific, Singapore 1986) pp. 273-290.

\noindent
16. Hadjiivanov, L.K. , private communication.

\noindent
17. Floreanini,R.,Spiridonov,V.P. and Vinet, L., Comm.Math.Phys.
{\bf 137}, 149 (1991).

\noindent
18. Khoroshkin, S.M. and Tolstoy, V.N., Comm.Math.Phys, {\bf 141}, 599 (1991).

\noindent
19. Palev, T.D. and Tolstoy, V.N., Comm.Math.Phys, {\bf 141}, 549 (1991).

\noindent
20. Scheunert, M., Nahm, W. and  Rittenberg, V., J.Math.Phys.
{\bf 18}, 155 (1977).

\noindent
21. Ky, N.A., Palev, T.D. and Stoilova, N.I., J.Math.Phys. {\bf 33},
1841 (1992).

\end